\documentclass[twocolumn]{revtex4}

\def\be{\begin{equation}}
\def\ee{\end{equation}}
\def\la{\label}
\def\bea{\begin{eqnarray}}
\def\eea{\end{eqnarray}}
\def\non{\nonumber}
\def\ci{\cite}
\def\la{\label}
\def\bib{\bibitem}

\def\lm{\lambda}

\def\le{\left}
\def\ri{\right}

\def\vp{\varphi}
\def\Omvp{\Omega_\vp}

\def\Om{\Omega}

\def\rp{\rho_\phi}

\def\wp{w_\phi}

\def\s8{\sigma_8}

\def\fr{\frac}
\def\pp{\partial}
\def\pu{\pp_\mu}
\def\pU{\pp^\mu}

\def\non{\nonumber}

\def\r{\rho}
\def\rvp{\rho_\vp}

\def\rvpbd{\rho_{\vp BD}}

\def\rp{\rho_\phi}

\def\wp{w_\phi}

\def\Ompbd{\Omega_{\phi BD}}
\def\Omvpbd{\Omega_{\vp BD}}

\def\G{\Gamma}

\def\vin{V_{int}}

\def\vip{V_{int,\vp}}

\def\mmp{m^2_{\phi }}
\def\mmpo{m^2_{\phi 0}}

\def\mmvp{m^2_{\vp }}

\usepackage{graphicx}

\begin{document}

\title{Inflation-Dark Energy unified through Quantum Regeneration}

\author{A. de la Macorra and F. Briscese }
\affiliation{Instituto de F\'{\i}sica, Universidad Nacional Autonoma de Mexico, Apdo. Postal 20-364,
01000 M\'exico D.F., M\'exico\\
Part of the Collaboration Instituto Avanzado de Cosmologia}

\begin{abstract}

We study a scalar $\phi$ field that unifies  inflation and dark
energy with a long period of a hot decelerating universe in between
these two stages of inflation. A key feature is that the transition
between the intermediate decelerated phase to the dark energy phase
is related to a quantum regeneration of the scalar field $\phi$
instead to purely classical dynamics. The interaction
$\vin(\phi,\vp)$ between $\phi$ and a second scalar field $\vp$
allows not only for $\phi$ to decay  at a high energy $E_I$,  even
though $\phi$ does not oscillate around the minimum of its
potential, but it also regenerates $\phi$ at an energy $E_{BD}\simeq
E_I^2$ close to present time. It  predicts extra relativistic energy
and an interacting dark energy which may account for a $w<-1$. In
the model presented here we have $E_I\simeq 100 \,TeV$ and
$E_{BD}\simeq 1\, eV$. The low value of the back decay energy
$E_{BD}$ gives an explanation of the coincidence problem.

\end{abstract}

\pacs{}

\maketitle

In the last few years the existence of two stages of
accelerating the universe have been established. The first stage is
 {\it inflation}  \cite{inflationarycosmology} needed to explain the
homogeneity and isotropy of our universe and the second  stage
corresponds to the dark energy "DE" which dominates the universe at
present time  \cite{wmap}.

Inflation is  associated with a scalar field, the "inflaton", and
the energy scale at which inflation occurs is typically of the
order of $E_{I}= 10^{16} GeV$ \cite{inflationarycosmology}  but it
is possible to have consistent inflationary models with $E_I$ as
low as  $ 10 MeV$\cite{lowinflation}. On the other hand the energy
scale at which DE  becomes relevant is much smaller than for
inflation, namely $E_o\simeq 10^{-3} eV$, and its
nature is not known. The present time data are consistent with a
cosmological constant  but perhaps the most appealing candidate
for the DE is that of a scalar field \ci{Q}-\ci{mio} which
interacts   weakly with  standard model "SM" particles.

We will assume that DE and inflation are given in terms of the
same scalar field and we will call this field the "uniton" $\phi$,
from inflaton-dark energy unification.
A single scalar field can easily give inflation and DE if
the potential is flat at high and low energies (present
time)\cite{scalarunification}. However, most of the time our
universe was decelerating and was not dominated by the uniton
field but by radiation and later by matter. Therefore, any
realistic model must not only explain the two stages of inflation
but also allow for a long period of decelerating universe. This
long period of deceleration is not easy to achieve in the context
of inflation-dark energy unification.
In this letter we study a class of models that achieve these
results.

To reheat the universe and
obtain a long period of deceleration  we couple the uniton
to a scalar $\vp$.  After inflation the uniton will decay into
$\vp$, reheating the universe,  while at low energies (close to
present time) $\vp$ will decay back  and regenerate the uniton.
We use for this second decay the same interaction term $\vin$ as in the
first decay. We
point out  that the appearance of DE is via quantum decay and
not  classical evolution.

Typically the inflaton decays while it oscillates around
the minimum of its (quadratic) potential
\ci{inflationarycosmology}. If the inflaton decay is not complete
then the remaining energy density of the inflaton redshifts  as
matter at late times. The amount of residual energy density
must be fine tuned if one wants to be interpreted as dark matter.
However, in our case the uniton can no longer have a
minimum at a finite value for $\phi$ \ci{mio} since it must
accelerate the universe at late times. So the decay of $\phi$ must
happen while rolling down its potential. Since   the uniton  inflates at an early and a late time
the potential $V$ must be flat at high and low energies,
we can take  $\phi<-1$ for inflation and $\phi>1$ for
dark energy, and then the uniton   evolves
through the region $\phi=0$ with a non vanishing potential
$V\neq 0$. The conditions for instant preheating \ci{ip} are then easily met
and we have an efficient  decay. Of course we want to reheat the universe with particles of the
standard  model "SM". To achieve this we couple $\vp$ to SM particles at high energies, the same
energy as the decay of $\phi$, where all particles of the SM and
$\vp$ are relativistic.  Thermal equilibrium "TE" will be maintained
as long as $\vp$ and the SM particles remain  relativistic.

The quantum regeneration scenario for DE, presented here,  has some
interesting generic properties and can be observationally or
experimentally tested. The explicit form of the uniton potential
$V(\phi)$ is not important as long as if inflates the universe at an
early and late epoch and the uniton field evolves a through region
(e.g. $|\phi|\ll 1$) where the conditions for instant preheating are
met. The field $\vp$  remains relativistic until present time and
therefore  we have more  relativistic energy density given by
$\Omvp=\Omega_{r} g_\vp/g_r$ with $g_\vp=1,g_r=g_\vp+g_{SM}$ the
relativistic degrees of freedom.   This extra $\Omvp$  is favored by
the cosmological data \ci{rel.d} and since the interaction between
DE and $\vp$ remains   at low energies it can also have
phenomenological consequence in structure formation and the
evolution of DE. In fact, an interacting DE has been proposed to
explain a $w$ smaller than -1 for DE \ci{ide}. In models where
inflation and reheating takes place at a low energy, as in the model
presented here with $E\leq E_I=O(100) \,TeV$, the temperature is
large to produce all SM particles but low enough so that $\vp$ could
be produced at LHC.  Of course we should be careful not to
contradict present day constrained  from   charged particles
\ci{Q.C} or a long range force
\ci{G.C}.\\
{\bf General Framework}--
Our starting point is a flat FRW universe with the uniton
$\phi$, a second $\vp$ scalar field and the SM. We take the
lagrangian $L=L_{SM}+\widetilde{L}$, where $L_{SM}$ is the SM
lagrangian, $\widetilde{L}=\fr{1}{2}\pu \phi \pU \phi +
\fr{1}{2}\pu \vp \pU \vp -V(\phi)-B(\vp)- V_{int}(\phi,\vp,SM)$. The uniton potential
is $V(\phi)$ while $\vin $ is the complete
interaction potential. The potential $B(\vp)$ may be
required to stabilize $\vp$, e.g. $B(\vp)=\lm\vp^4$ for $\vin$ in
eq.(\ref{vin}).
The requirement  for $V$ is that it satisfies the
slow roll conditions $|V'/V|<1, |V''/V|<1$, where a prime denotes
derivative w.r.t. to $\phi$, at the inflation epoch and at present
time for DE. We also take $V$ such that $\phi$ evolves  through
regions where instant preheating is possible, e.g.  $V(\phi=0)\neq
0$. The interaction term $\vin$ has two important consequences. On
the classical level it couples the differential equations of
$\phi$ and $\vp$ through derivatives of $\vin$ while at a quantum
level it allows for a particle decay.

We  define the energy
density and pressure for the field $\phi$ as
 $\rp =\fr{1}{2}\dot\phi^2+  V$,
$ p_\phi =\fr{1}{2}\dot\phi^2 -V$ and $\r_\vp =\fr{1}{2}\dot\vp^2+
B+ \vin$, $ p_\vp =\fr{1}{2}\dot\vp^2 - B- \vin $ for $\vp$. The total
energy density and pressure are then given by $\rho=\rp+\rvp,
p=p_\phi+p_\vp$. The classical evolution of $\phi$ and $\vp$ \ci{csf} is
given by the equations of motion, $\ddot\phi+3H\dot\phi+V'+ \vin'
=0$, $ \ddot\vp+3H\dot\vp+B_\vp+ \vip=0$, with $H^2= \rho/3$ and
$8\pi G\equiv 1$.
In a non expanding universe the number density $n=N/\textrm{Vol}$,
where $N$ is the total number of particles
and $\textrm{Vol}$ the volume, evolves as $n(t)=n_i e^{-\G (t-ti)}$
where $\G$ is the transition (constant) rate.
The   differential transition rate is given by \ci{G}
$
d\G= \textrm{Vol}(2\pi)^4
|M_{ab}|^2 \delta^4 (PI-PF)\Pi_a \fr{1}{2E_a \textrm{Vol}} \Pi_b
\fr{d^3p_b}{2E_b (2\pi)^3}
$
where $PI(PF)$ is the initial (final)
momentum, $\textrm{Vol}$ is the volume (normalized to one particle
per volume) and  $M_{ab}\equiv \langle b|M|a\rangle$ is the transition amplitude.
The conservation of energy-momentum requires that initial and final energies
are equal, $E_i=E_f$ and $p_i=p_f$.
In a process of $a$ identical initial particles with energy $E_a$
and mass $m_a$ and a final state consisting of $b$ particles with
the same energy $E_b$ and mass $m_b$ so that $E_i=a\,E_a =bE_b=E_f$
 differential transition rate is
\be\la{G}
\G=c_{ab}
|M_{ab}|^2n_a^{a-1}p_b^{b-1}E_a^{b-a-3}
\ee
$c_{ab}=(a/b)^{b-2}2^{(1-a-b)}(2\pi)^{3-2b}/a $.
In the limit where the decaying particle is non-relativistic
with $E_a\simeq m_a\gg m_b, p_b\simeq E_b$ then eq.(\ref{G}) becomes
\be\la{Gnr}
\G= c_{ab} |M_{ab}|^2 n_a^{a-1} E_a^{2b-a-4}= c_{ab} |M_{ab}|^2 n_a^{a-1} m_a^{2b-a-4}
\ee
On the other hand
if all particles involved are relativistic and in TE then eq.(\ref{G})
with $n_a= c_nT^3, c_n=g_a\zeta(3)/\pi^2$ and $E_a=T$ is
\be\la{Gr}
\G=\widetilde{c}_{ab}|M_{ab}|^2  E_a^{2(b+a)-7},
\ee
$\widetilde{c}_{ab}=c_{ab}c_n^{a-1}$.
In quantum field theory it is common to take the interaction between
two scalar fields as power laws with $\vin= g\, \phi^m\vp^n/2$
with $m>0, n>0$ and $n+m\leq 4$.
However, the potential for DE is in general
a non renormalizable potential and has a more complicated expression
such as an exponential  $e^{-\alpha \phi }$ or an inverse power $1/\phi^\alpha$.
Therefore, we will consider a generic interaction potential $\vin(\phi,\vp)$.
The quantum states in field theory  are perturbations around
the minimum of the potential, however, since a scalar field that
is cosmologically evolving has not reached its lowest energy
state, we must  expand $\phi$ around its classical average
$\phi_0(t)$ at any given time,
 $\phi(t)=\phi_0(t)+\delta \phi(t)$,
and it is the fluctuation $\delta \phi$  that gives the quantum
state. An expansion around a stable point of the potential $V$, as
in a quadratic potential, the creation of particles are not
energetically favored. However when the perturbations are unstable
the creation of particles is energetically favored. Let us assume
an interaction term $\vin=g h(\phi) \vp^n$ with $h$ not
necessarily a positive power law function of $\phi$.  If we expand
$h$ in a Taylor series around $ \phi_0(t)$ the interaction term
$\vin$ gives an effective coupling $\la{vi2} \vin\simeq g h_0\vp^n
+ g h'_0\, \delta\phi\;\vp^n +\fr{1}{2}g
h''_0\,\delta\phi^2\;\vp^n +... $ between   $a$ quantum fields
$\delta\phi$ with $1\leq  a$ and $b$ quantum fields $\delta\vp$
with $1\leq b\leq n$, after expanding $\vp=\varphi_0+\delta\vp$.
Since $\phi$  is dynamically evolving, the expansion point
$\phi_0$ and all the couplings $h_0, h'_{0},...$ are functions of
time. Considering a polynomial interaction we can determine the process of an
initial state of a-particles going into a final state of
b-particles. The transition amplitude is
\be\la{mab}
 M_{ab}=\fr{1}{a!b!}\fr{d^a}{d\phi^a}\fr{ d^b \vin }{d\vp^b}
\ee
and the total transition rate  is then given by
$\G=\Sigma_{a,b}\G_{ab} $ where $a$ takes the value from $1\leq
a\leq a_{max}$ and $1\leq b \leq b_{max}$. Clearly the total
transition rate $\G$ will be dominated by the largest $\G_{ab}$.
If we take a polynomial potential $\vin(\phi,\vp)=g\,\phi^m\vp^n $
with arbitrary   values of $m,n $ and use eq.(\ref{mab})   we have
$M_{ab}=\fr{m!n!}{a!(m-a)!b!(n-b)!}\;g\phi^{m-a}\vp^{n-b}$ and
eq.(\ref{G}) becomes
 $\G_{ab} =\G_{12} \G_i^{a-1}\G_f^{b-2}$ with
$\G_0 \equiv   c_0g^2\phi^{2(m-1)} \vp^{2(n-2)} / m_\phi,\,
\G_i\equiv   n_\phi/\phi^2 m_\phi,\, \G_f\equiv
m_\phi^2/\varphi^2$ and $ c_0=(\fr{m!n!}{a!(m-a)!b!(n-b)! })^2
c_{ab}$. The quantity $\G_{12}$ corresponds to the decay with
$a=1, b=2$, $\G_i$ gives the contribution from  a larger number
($a>1$) of initial decaying particles $\phi$ while $\G_f$
corresponds to a larger number ($b>2$) of final product particles
$\vp$.   We clearly  see that    if $\G_i=n_\phi/\phi^2 m_\phi>1$ then a
large value of  $"a"$ gives a bigger $\G_{ab}$ or   if
$\G_f=m_\phi^2/\vp^2 > 1$  when  $"b"$ takes
its maximum value $b=n$.\\
{\bf Interaction term}-- We will now choose an interaction term
which allows for $\phi$ to decay into a relativistic scalar field
$\vp$ at a high energy $E_I$. This field $\vp$ is coupled to SM
particles denoted by $\chi$ and $\psi$. As soon as $\vp$ is
produced it decays into $\chi$ and $\psi$, which are also
relativistic at the energy $E_I$. As long as $\vp$, $\chi$ or
$\psi$ are  relativistic they remain in TE. We will assume that
$\vp$ remains relativistic while it is couple to the SM (otherwise
the number density $n_\vp$ would be exponentially suppressed and
the uniton would not be regenerated). Finally, the $\vp$ will
regenerate $\phi$ at a late time when the universe energy is given
by $E_{BD}$, with $E_I\gg E_{BD}> E_o=10^{-3} eV$. From
eq.(\ref{Gr}) we see that if want to have a late time regeneration
$E\ll 1$  the coupling between $\vp$ and $\phi$, which  are
relativistic, must have $a+b< 4.5$ so that the exponent of $E$ in  $h^2
E^{2(a+b)-9}\geq \G/H$ is negative, and we have used $H\approx\sqrt{\rho}\sim E^2$.
We take then the simplest interaction
potential is
\be\la{vin}
\vin(\phi,\vp,SM)=g\,\phi\,\vp^3+h\,\vp^2\chi^2+
\widetilde{h}\,\vp^2\bar\psi\psi.
\ee
The first term gives rise to
the uniton decay into $\vp$ at a high energy and a late time
regeneration via $\vp$ decay. The second and third terms are the
coupling of   $\vp$  with SM particles $\chi,\psi$   allowing for
reheating our universe.  All other SM particles will  be produced
by $\chi,\psi$. If the  fields $\chi,\psi$  acquire a large mass
then $\vp$ will no longer be coupled at $T< m_\chi$ since below
this  temperature $n_\chi, n_\psi$ are exponentially suppressed
and $\G/H$ will be smaller than one. However, the $\vp$
temperature will still redshift as $T\sim 1/a(t)$ since it is
relativistic.

Let us now describe the three different decay processes, the
uniton decay, the SM reheating and the back regeneration of the
uniton. \\
{\bf Inflaton Decay}-- In the decay of
$\phi$ two different and complementary scenarios take place. On
the one hand we have $\G/H\gg 1$ giving an exponentially
suppressed $\rp$ after the decay. On the other hand, the
conditions for a non adiabatic process  and instant preheating are
met, i.e. we have $\dot m_\vp/\mmvp >1$, since the potential
$V(\phi)\neq 0$ while $\phi$ roles down its potential around the
value $\phi=0$ and the mass of $\vp$, $\mmvp=6g\phi\vp$, vanishes.
Taking into account these facts we expect an efficient decay of
$\phi$ into $\vp$. Furthermore, since $\vp$ is coupled to the SM
it will decay into SM particles and the resulting energy density
$\rp$ will essentially vanish giving rise to a relativistic
$\rho_R=\rho_{SM}+\rvp$. If inflation ends with a small value of
$\phi$ than the decay and reheating of the universe will take
place immediately afterwards at the  energy scale $E_I$. After
inflation the field $\phi$ is non-relativistic, in the comoving
frame its momentum is negligible compared to its mass, and we have
$\mmp> V\simeq \rp$. Form eq.(\ref{Gnr}) with $a=1, b=3$ and
$E_a=E_\phi=m_\phi$ we have
\be\la{d1}
 \G=\fr{g^2 m_\phi}{192\pi^3},\;\;H=\sqrt{\fr{\rp}{3}}, \hspace{.5cm}
\fr{\G}{H}=\fr{g^2  }{192\pi^3}\sqrt{\fr{ 3\mmp}{\rp}}.
\ee
We see form eq.(\ref{d1}) for as long as $g^4\mmvp\gg  \rp$ the field
$\phi$ will decay into $\vp$. If $\G/H\gg 1$ then we will have an
efficient decay and $\rp$ will be exponentially small. Now, the
instant preheating takes place the non adiabaticity condition  \ci{ip}
\be\la{c.m.}
 \le|\fr{\dot
m_\vp}{\mmvp}\ri|=\le|\le(\fr{\dot\phi}{\phi}+
\fr{\dot\vp}{\vp}\ri)\fr{1}{2m_\vp}\ri|\geq
\le|\fr{V^{1/2}}{m_\vp\phi}\ri|\gg 1
\ee
where we have used
$\dot\phi^2=2(1+\wp)/(1-\wp) V$. After inflation the energy
density $\rp$ redshifts with an equation of state $\wp\neq -1$
and $m_\vp \phi\approx 0$ for $\phi\approx0$ giving $
|\dot m_\vp/\mmvp|\gg 1$ in eq.(\ref{c.m.}). \\
{\bf Universe Reheating}-- The
reheating of the universe takes place via a process
$\vp+\vp\leftrightarrow \chi+\chi$ (or
$\vp+\vp\leftrightarrow\psi+\psi$)  with a cross section for
relativistic particles $\sigma = h^2/E^2$ (we take the same
strength for   the $\chi $ and $\psi$) and an interaction rate
\be\la{dr}
 \G=\fr{h^2 E}{32\pi^3},\;\; H=\sqrt{\fr{\rho_r}{3\Omega_r}} \equiv c_H E^2,
\hspace{.5cm} \fr{\G}{H}=\fr{c_R\,  h^2  }{E}
\ee
with $T=E,\,\widetilde{c}_{22}\simeq 1/32\pi^3$,$ c^2_H\equiv g_r\pi^2/90\Omega_r$  and
 $c_R\equiv c(E_R),c(E)\equiv (c_H 32\pi^3)^{-1}$.
For $E> 10^2 GeV$ we have $g_r\simeq 106, \Omega_r\simeq 1$ and
$c_R\simeq 10^{-3}$. Clearly eq.(\ref{dr})
maintains a TE for $E\leq E_R\equiv c_R h^2$. A good choice of $h$ is then
$h^2\simeq E_I$ so that the interaction takes place
at  $E_R=T_R\simeq E_I$. The amount of $\rvp$ can then be easily
determined and it is $\Omvp= \Om_{r}/g_{r}$. In terms of $\Delta N_\nu$, extra
neutrinos degrees of freedom, we have $ \Delta
N_\nu=(8/7)(g_\vp/g_\nu) (T/T_\nu)^4$ with   $ \Delta N_\nu=2.2\,
(0.57)$ for $T=T_\gamma (T_\nu)$ (if $\vp$ decouples at a higher
energy than the neutrinos then $ \Delta N_\nu<0.57$). A central
value of $0.5<\Delta N_\nu<2.1$ is favored by the cosmological
data  \ci{rel.d}.\\
{\bf Back Decay and quantum regeneration--} Now, let us see the
case for the back decay and quantum regeneration of the uniton
field $\phi$. This process will take place at late time and low
energies $E=E_{BD}$. Therefore the classical potential $V\ll
E_{BD}$ and  $m_\phi\ll E_{BD}$. So the uniton will be
relativistic and from eq.(\ref{vin}) the process
$\vp+\vp\rightarrow \vp+\phi$ gives
\be\la{dbd}
 \G=\fr{g^2 E}{32\pi^3},\;\; H=\sqrt{\fr{\rho_r}{3\Omega_r}} =c_H E^2,
\hspace{.5cm} \fr{\G}{H}=\fr{c_{BD}\, g^2  }{E} \ee For  low energy,
$E\ll MeV$, we have $g_r\simeq 5$ and $c_{BD}\equiv c(E_{BD})\simeq
10^{-3} \sqrt{\Omega_r} $. The process takes place for $E\leq
E_{BD}\equiv c_{BD} g^2 $. An interesting choice  is $g\simeq  E_I$,
which gives $E_{BD}\ll E_I$. With this choice we reduce the number
of free parameters and we relate the energy scale of the back decay
to that of the end of inflation
\bea\la{ghe}
E_{R}\equiv c_R g^2\approx E_I^2,&&     E_{BD}\equiv c_{BD} h^2 \approx E_I,\non\\
 g=h^2&=&q  E_I
\eea
with $q$
a proportionality constant. The fine structure constant of these
interactions are $\alpha_I\equiv h^2/4\pi=E_I/4\pi$ and
$\alpha_{BD}\equiv g^2/4\pi=E^2_I/4\pi$ which for $E_I=100\,TeV$
gives $\alpha_I=10^{-14}, \alpha_{BD}=10^{-27}$ to be compared with
$\alpha_{em}=1/137$, the electromagnetic fine structure constant.
The constraint on light particles coupled to electrons from
astrophysical considerations is $\alpha < 0.5\; 10^{-26}$ \ci{Q.C}
or to baryons from a long  range force \ci{G.C} imply that the SM
field $\chi$ and $\psi$  must be a neutral particles
such as neutrino. \\
{\bf Uniton potential  ${\bf V(\phi)}$} -- The choice of $V$ is not
essential in this class of models as long as it is flat at a high
energy and late time to give an accelerating universe and that the
evolution of the field $\phi$ go through values where $V\neq 0$ with
$m_\vp=0$ so that $|\dot m_\vp/\mmvp|\gg 1$. However, to be more
specific we  present as an example the potential \be\la{v}
V(\phi)=\frac{V_I}{2}\le(1-\fr{2}{\pi} \arctan[k\phi]\ri) \ee with
$V_I, k$ constant parameters. The potential in eq.(\ref{v})  can be
motivated by the interaction $AB\rightarrow C \rightarrow A'B'$,
with the exchange of a scalar particle with propagator
$1/(E_c^2-p_c^2-m_c^2)$. The Yukawa potential $V_Y\propto
e^{-m\,r}/r$ is obtained as the fourier transformation for
$E_c\simeq 0$ while our potential corresponds to zero momentum with
$p_c=0$ and $E_c=m_A-m_B$. Integrating the propagator with imaginary
energy $E_c=i\widetilde{E}_c$ we get a potential $
V_s\propto\int_{-\infty}^{\infty}  dE_c/(E_c^2-m_c^2)=-i\pi/m_c $
and an  euclidian action $S_E=-iS\propto -iV_s\propto-\pi/m_c $.
This $S_E$ gives an exponentially suppressed transition rate
connecting a maximum (e.g. at $V_I$) to a minimum (e.g. $V=0$), i.e.
a sphaleron configuration. If we take the integration limits as
$E_{max}=m_A-m_B=\phi$ we have $V\propto\int_{-\phi}^{\phi}
d\widetilde{E}_c/(\widetilde{E}_c^2+m_c^2)=2 \arctan[\phi/m_c])/m_c$
corresponding to $V$ in eq.(\ref{v}). We could identify the energy
scale $ V_I$ as the ultraviolet cutoff scale above which susy (or
another symmetry) gives a vanishing $V$.

We constrain the values of $V_I,
k$ in eq.(\ref{v}) by demanding that during inflation $c\delta \rho/\rho=
V^{3/2}/V'=5.2\times 10^{-4}, c=\sqrt{75\pi^2}$ and at present time the DE density
$ V(\phi_o=1)\simeq V_o=(2\times 10^{-3} eV)^4$. The inflation, reheating and
back decay scales, using eq.(\ref{ghe}) with $q=10$,  are
\be
E_I \simeq  100\,TeV,\;\; E_R \simeq 1\, TeV,\;\;
 E_{BD} \simeq  1 \,eV
 \ee
The  scale $E_I$ is very interesting
since it is on the upper limit of susy.  This inflationary scale $E_I$ is
low compared to the standard $10^{16} GeV$ but it is large enough to have a
reheating temperature  to produce all SM particles and it is
within the phenomenological range at LHC. Moreover, it is phenomenological welcome
\cite{lowinflation} and since it is low scale one does not have gravitino overabundance
problems and it has a spectral index $n_s=0.97$. It would be
interesting to relate the coupling $ E_{BD}=1\, eV$ to
the neutrino mass \ci{fabio}.

The potential in eq.(\ref{v}) has
$V'=-V_Ik/[\pi(1+k^2\phi^2)]$,
$\mmpo=V''=2V_Ik^3\phi/[\pi(1+k^2\phi^2)^2]$ and the limits
$V(-\infty)=V_I, V(0)=V_I/2$ and $V(\infty)=0$.
If we take $k\gg
1$ (as will be needed later on) then the potential $V$ satisfies
the slow roll conditions and accelerates the universe for $\phi<
-k^{-1/3}$ and for $\phi \geq 1$.
During the last 60 e-folds
of inflation we have $ k^{-1/3}\leq -\phi \ll 1$ and the potential
can be approximated by  $V\simeq V_I$,
 $V'\simeq -V_I/ k\phi^2\simeq -V_I k^{-1/3}$ and
$V^{3/2}/V'\simeq V_I^{1/2} k^{1/3}$ while for $\phi\simeq 1$ we
have $V(\phi_o\simeq 1)=V_o\simeq V_I/k$. We then obtain the
values $ k\simeq 10^{66}\gg 1$ and $ V_I\simeq 10^{-53}\ll 1$. The
dimension of $V_I$ is $[V_I]=[E^4]$ so we obtain an inflationary
epoch $E_I=V_I^{1/4}\simeq 100 TeV$ while $[k]=[1/E]$ and
$1/k=O( E_I (V_I/m_{pl}^4))=O(E_I^5)$. Using eq.(\ref{d1}) we have
the $\phi$ decay rate $ \G  \approx g^2 m_\phi = E_I^2 V_I^{1/2}k $
and
\be
\fr{\G}{H} \approx  g^2 \sqrt{\fr{\mmp}{\rp}} \simeq
E_I^2 k=10^{40}\gg 1
\ee
and eq.(\ref{c.m.}) is also satisfied.
Therefore we have an efficient decay and the field $\phi$ ceases
to exist until it is regenerated at late time by the back decay
process. The universe will therefore be in a decelerating phase
for a long period of time, from reheating at $E_I\simeq 100\,TeV$ to
$\vp$ decay at $E_{BD}=E_I^2=1\, eV$ (c.f. eq.(\ref{dbd})) when
$\vp$ starts to regenerate $\phi$ giving $\Ompbd=\Omvpbd$ with
$\rvpbd\simeq E_{BD}^4=E_I^8$. For  $V\approx\rp\approx \rvpbd $
we have
 $\phi>1/kV_I\simeq 10^{-12}$, using
$V\simeq  V_I/k\phi$ (valid for $\phi k\geq 1$). Once $\phi$ is
regenerated it will grow and its potential will start dominating the
universe with  $\phi=O(1)$ for $V\approx V_o$, independent of its
initial conditions (tracker behavior). The slow roll conditions are
satisfied and the universe will enter an acceleration period or DE
domination. This late time decay gives
an understanding why DE appears at such a late time.\\
{\bf Summary and conclusions}-- We have presented a model
where inflation at and dark energy can be achieved via a single
scalar field $\phi$, the uniton. In order to have a long period
of hot and decelerating universe we couple $\phi$ to another field
$\vp$.   This coupling allows for the reheating of our
universe right after inflation at $E_I=100\, TeV$, a reheating scale $E_R=1\,TeV\simeq E_I$
 and a late time regeneration of
$\phi$ at a low energy $E_{BD}=1\,eV\simeq E_I^2 $ which could be related to the
neutrino mass.
More relativistic energy and a $w<-1$  are phenomenological favored \ci{rel.d}
and our model can explain both since we have an interacting
dark energy, which may account for a $w<-1$, and predicts
the existence of $\Omvp$.
The universe  is dominated by $\rp$  at high energy $\rp>
E_I^4=(100\, TeV)^4$ and low energy $\rp\ll \rho_{BD}=(1\, eV)^4$ while
radiation or matter dominates the universe otherwise. We
stress the fact that the quantum regeneration of the uniton
drives the transition between the decelerating
universe to the dark energy phase, it is not longer
classical  but it is essentially due to quantum effects and
the low value of $E_{BD}$
explains the coincidence problem.\\
\noindent{\bf Acknowledgment}-- A.M.would like to thank G. German for
discussion at an early stage of this manuscript. We thank for
partial support Conacyt project 45178-F, IAC-Conacyt Project
and HELEN network.

\end{document}